\documentclass[conference]{IEEEtran}
\IEEEoverridecommandlockouts
\usepackage{cite}
\usepackage{amsmath,amssymb,amsfonts}
\usepackage{algorithmic}
\usepackage{graphicx}
\usepackage{enumitem}
\usepackage{textcomp}
\usepackage{xcolor}
\def\BibTeX{{\rm B\kern-.05em{\sc i\kern-.025em b}\kern-.08em
    T\kern-.1667em\lower.7ex\hbox{E}\kern-.125emX}}

\usepackage{subfig}
\usepackage{multirow}
\usepackage{tabularx}
\usepackage{array}
\usepackage{graphicx} % This package is essential for including images

\usepackage{makecell} % For \Cline
\newcolumntype{I}{!{\vrule width 1.5pt}}

\usepackage[framemethod=tikz]{mdframed}
\mdfdefinestyle{mpdframe}{
    frametitlebackgroundcolor   =black!15,
    frametitlerule              =true,
    roundcorner                 =5pt,
    middlelinewidth             =1pt,
    innermargin                 =0.3cm,
    outermargin                 =0.3cm,
    innerleftmargin             =0.3cm,
    innerrightmargin            =0.3cm,
    innertopmargin              =0.5cm,
    innerbottommargin           =0.5cm
}

\newcommand{\eg}{\textit{e}.\textit{g}., }

\usepackage[numbers]{natbib}

\begin{document}

\title{Investigating Developers' Preferences for Learning and Issue Resolution Resources in the ChatGPT Era
}

\author{
\IEEEauthorblockN{Ahmad Tayeb\textsuperscript{\textdagger}, Mohammad Alahmadi\textsuperscript{*}, Elham Tajik\textsuperscript{\textdagger}, Sonia Haiduc\textsuperscript{\textdagger}} 
\IEEEauthorblockA{\textsuperscript{\textdagger}\textit{Florida State University, United States}\\
\textsuperscript{*}\textit{University of Jeddah, Saudi Arabia}} 
\IEEEauthorblockA{\textsuperscript{\textdagger}\{atayeb2, et22e, shaiduc\}@fsu.edu, \textsuperscript{*}mdalahmadi@uj.edu.sa}
}

\maketitle

\begingroup\renewcommand\thefootnote{\textdagger}
\footnotetext{Ahmad Tayeb is also with King Abdulaziz University, Saudi Arabia.}
\endgroup

\maketitle

\begin{abstract}
The landscape of software developer learning resources has continuously evolved, with recent trends favoring engaging formats like video tutorials. The emergence of Large Language Models (LLMs) like ChatGPT presents a new learning paradigm. While existing research explores the potential of LLMs in software development and education, their impact on developers' learning and solution-seeking behavior remains unexplored. To address this gap, we conducted a survey targeting software developers and computer science students, gathering 341 responses, of which 268 were completed and analyzed. This study investigates how AI chatbots like ChatGPT have influenced developers' learning preferences when acquiring new skills, exploring technologies, and resolving programming issues. Through quantitative and qualitative analysis, we explore whether AI tools supplement or replace traditional learning resources such as video tutorials, written tutorials, and Q\&A forums. Our findings reveal a nuanced view: while video tutorials continue to be highly preferred for their comprehensive coverage, a significant number of respondents view AI chatbots as potential replacements for written tutorials, underscoring a shift towards more interactive and personalized learning experiences. Additionally, AI chatbots are increasingly considered valuable supplements to video tutorials, indicating their growing role in the developers' learning resources. These insights offer valuable directions for educators and the software development community by shedding light on the evolving preferences toward learning resources in the era of ChatGPT.

% The landscape of software developer learning resources has continuously evolved, with recent trends favoring engaging formats like video tutorials. The emergence of Large Language Models (LLMs) like ChatGPT presents a new learning paradigm. While existing research explores the potential of LLMs in software development and education, their impact on developers' learning and solution-seeking behavior remains unexplored. We address this gap by conducting a survey targeting software developers and computer science students, investigating how AI chatbots like ChatGPT have influenced developers' learning preferences when acquiring new skills, exploring technologies, and resolving programming issues.
% More specifically, through quantitative and qualitative analysis, we study whether AI tools supplement or replace traditional learning resources such as video tutorials, written tutorials, Q\&A forums, etc. We also investigate the perceived advantages and disadvantages of AI chatbots for learning programming topics and solving programming issues. Our findings offer valuable insights for educators and the software development community by investigating how developers’ preferences toward various learning resources have evolved in the era of ChatGPT.
\end{abstract}

\setlist[itemize]{leftmargin=*}

\begin{IEEEkeywords}
AI-driven learning tools, ChatGPT, programming learning preferences, programming issue resolution, video tutorials, written tutorials, Q\&A forums, developer learning resources.
\end{IEEEkeywords}

\section{Introduction}\label{sec:introduction}
The array of learning resources available to software developers has always been wide-ranging, extending from traditional textbooks to dynamic platforms such as Stack Overflow and YouTube. The increasing popularity of video tutorials, as discussed in the works of \citet{macleod2015code} and \citet{escobar2019survey}, indicates a shift towards more engaging and immersive learning experiences. Complementing this trend, \citet{kafer2017best} provide an insightful comparison between video and text tutorials, examining their efficiency and effectiveness. Their study, which involved converting a video tutorial into a text format and testing it among undergraduate students, found nuanced differences in learning efficiency. They observed that while text tutorials were completed faster, video tutorials led to the quicker application of learned content, and learners preferred video tutorials for new content when both formats were available but relied on text tutorials for quick information retrieval \cite{kafer2017best}.

The emergence of Large Language Models (LLMs) like ChatGPT has influenced software engineering and computer science education. Recent research highlights their application in tasks like code completion, debugging, and improving software security reveals the potential of these models to enhance software development practices \cite{xu2022systematic, chaaben2022using, sobania_analysis_2023, kang2022large, akli2022predicting, lyu2023chronos, le2023log, nashid2023retrieval, siddiq2023zero}. Furthermore, studies in the field of computer science education have shown the effectiveness and implications of using ChatGPT in education, from generating code solutions to participating in assessments and personalized learning support \cite{savelka2023can, zhai2023chatgpt, qureshi2023exploring, savelka2023large, becker2023programming, joshi2023chatgpt, jalil2023chatgpt, elsayed2023towards}. ChatGPT, a chatbot developed by OpenAI, uses artificial intelligence to generate text responses that mimic human conversation, including answering follow-up questions based on prior context.

Despite existing research on developers' learning resources, the specific impact of AI-driven chatbots (\eg ChatGPT ), particularly in shaping programming learning and solution-seeking to programming issues behaviors, remains underexplored. Our research aims to bridge this gap by investigating how developers' preferences towards various learning resources have evolved in the era of ChatGPT. We are particularly interested in understanding developers' preferred methods for acquiring new skills, exploring new technologies, and resolving programming-related issues. Additionally, we are interested in understanding the role of AI chatbots in relation to traditional learning resources such as video tutorials, written tutorials, and Q\&A forums. 

To investigate this, we conducted a survey targeting developers and computer science students, garnering a total of 341 responses, with 268 completed and analyzed. The survey assessed their current use of programming learning resources, their experiences with AI-driven tools like ChatGPT, and their overall preferences for learning programming and programming issue resolution resources. Using quantitative and qualitative data, we aim to explore trends in adopting traditional versus AI-driven learning resources. This can help us understand whether tools like ChatGPT supplement or replace traditional learning methods.

We found a growing integration of AI-driven chatbots into developers' learning resources. While video tutorials remain the primary choice for their visual and detailed explanations, AI chatbots are increasingly valued for their instant, personalized responses. Interestingly, a considerable number of respondents view chatbots as potential replacements for written tutorials, highlighting a shift towards more interactive learning methods. However, chatbots are seen as complementary to video tutorials, underscoring the context-dependent perception of AI tools as learning aids. Search engines and Q\&A forums maintain their relevance for quick information access and community insights. Our findings indicate a nuanced adoption of AI chatbots, either supplementing or replacing traditional resources based on the learning scenario, marking an evolution in developers' learning preferences in the ChatGPT era. 

Our study's findings are poised to offer valuable insights for educators and the broader software development community, facilitating adaptation to the changing landscape of learning resources. 

Below is a summary of our contributions:

\begin{itemize}
    \item We conducted a survey targeting developers and computer science students, collecting 341 responses, with 268 completed and analyzed, to understand their preferences towards various learning resources in the ChatGPT era. 
    \item We analyzed 1,520 open-ended survey responses, which required around 219 work hours, using open coding techniques to understand developers' perceptions of AI chatbots as programming learning and issue resolution resources. 
    \item We provide the replication package of our research that includes the survey, data collected, and detailed documentation of the analysis process, including identified themes and codes \cite{replicationPackage}.
\end{itemize}

\section{Related Work}\label{sec:related_work}
\subsection{Generative Artificial Intelligence in Software Engineering Tasks}

% Interest in the utilization of Large Language Models (LLMs) like ChatGPT in software engineering has grown significantly in recent years. In this subsection, we outline researches that focused on an empirical evaluation of ChatGPT in various software engineering tasks such as code completion, code debugging, and software security. 

Recent studies have highlighted the effectiveness of Large Language Models (LLMs) like ChatGPT in software engineering, particularly in code completion, debugging, and security. Researchers such as \citet{xu2022systematic} and \citet{chaaben2022using} have demonstrated the capabilities of LLMs in code analysis and development, emphasizing the advantages of few-shot learning for enhancing model completion tasks. Further investigations by \citet{sobania_analysis_2023} and \citet{kang2022large} revealed ChatGPT's potential in automatic bug fixing and few-shot testing, showcasing its efficiency in bug identification and reproduction. Other applications include the prediction of flaky test categories by \citet{akli2022predicting} and the introduction of Chronos for vulnerability detection by \citet{lyu2023chronos}, highlighting the broad applicability of LLMs in software engineering.

Additionally, the utility of ChatGPT in processing and analyzing log files was explored by \citet{le2023log}, indicating its role in system diagnostics. The importance of prompt selection in code-related tasks was examined by \citet{nashid2023retrieval}, while \citet{siddiq2023zero} demonstrated the use of zero-shot prompting for code complexity prediction with GitHub Copilot. 

Recent investigations by \citet{liang2024large} and \citet{xu2022ide} provide further insights into the usability and effectiveness of AI programming assistants. \citet{liang2024large} conducted a large-scale survey revealing that while developers appreciate these tools for reducing keystrokes and recalling syntax, they face challenges with generating code that meets specific requirements and controlling the tool's output. Similarly, \citet{xu2022ide} examined the use of machine learning methods for code generation within the PyCharm IDE, finding mixed results in productivity and code quality, but positive developer experiences and key areas for improvement.

In contrast to existing studies that primarily showcase the effectiveness of AI chatbots in performing a specific software engineering task, our research focuses on the extent to which developers opt to utilize AI chatbots as a means to expand their knowledge, particularly in comparison to alternative learning resources.

\subsection{Generative Artificial Intelligence in Computer Science Education}

The adoption of Generative AI, notably Generative Pre-trained Transformers (GPT) and ChatGPT, in computer science education represents a notable shift towards integrating advanced AI tools in teaching and learning practices. Studies such as those conducted by \citet{savelka2023can} and \citet{zhai2023chatgpt} have assessed the effectiveness of these AI models in programming courses and science learning, acknowledging their ability to generate accurate code and support personalized learning, albeit with certain limitations in handling complex tasks. Research by \citet{qureshi2023exploring} and \citet{savelka2023large} further explores ChatGPT's impact on learning outcomes, engagement, and its performance in multiple-choice assessments, revealing mixed results that highlight both its potential benefits and challenges.

The implications of AI-driven code-generation tools on educational practices have prompted discussions among educators and researchers. \citet{becker2023programming} and \citet{joshi2023chatgpt} emphasize the need for the education community to critically engage with these technologies, considering their unreliability in certain academic contexts. Moreover, \citet{jalil2023chatgpt} and \citet{elsayed2023towards} investigate strategies to leverage ChatGPT's capabilities in curriculum design and address concerns over academic integrity, suggesting methods to enhance learning engagement and develop AI-resistant educational materials. 

% \citet{becker2023programming} presented a position paper discussing the educational implications of AI-driven code-generation tools in introductory programming, emphasizing the urgency for the computing education community to actively engage with these emerging technologies. The research by \citet{joshi2023chatgpt} provided a critical view, demonstrating ChatGPT's unreliability in undergraduate computer science topics and advising caution in its academic use.

% \citet{jalil2023chatgpt}'s work further investigated ChatGPT's effectiveness in a software testing curriculum, finding a notable rate of correct responses and suggesting improved performance with specific prompting strategies. Besides, \citet{elsayed2023towards} tackled the challenge of AI tools potentially undermining academic integrity and learning quality, proposing an evolutionary approach to create AI-resistant questions, thereby fostering critical thinking and deeper learning engagement.

This body of research collectively underscores the burgeoning role of generative AI in computer science education, bringing to light its potential as a learning aid, the necessity for informed usage, and the imperative for educators to adapt and evolve their teaching methodologies in response to these advanced technologies. In our research, we further extend the knowledge in this area by investigating developers' learning preferences in the generative AI era. This exploration aims to understand how these emerging technologies are reshaping the landscape of educational resources and methodologies in computer science.

\subsection{Preferred Online Sources for Programmers}

Programmers' need for continuous learning to stay updated has led to the popularity of online resources like StackOverflow, YouTube, and ChatGPT, highlighting the importance of understanding their preferences for effective learning environments. Research has shown the growing use of YouTube for learning programming through screencasts, offering interactive experiences compared to text-based sources \citet{storey2014r, macleod2017documenting, pongnumkul2011pause, Khandwala2018codeMotion, macleod2015code, macleod2017documenting}, despite challenges in directly utilizing code from videos \cite{alahmadi2022vid2xml,alahmadi2022vid2meta,malkadi2023improving,alahmadi2024optimizing}. The introduction of ChatGPT may be shifting preferences towards prompt, text-based answers, suggesting a need to reevaluate developers' learning material choices.

Interviews with programmers by \citet{alghamdi2023websites} showed a strong preference for Stack Overflow for programming tasks, while \citet{arya2023programmers}'s research outlined strategies used by programmers in selecting online resources, shedding light on the transition from question formulation to resource identification. Studies have also focused on the role of search engines in code comprehension, debugging, and information acquisition \cite{gallardo2011kinds, xia2017developers, maalej2014comprehension, bai2020graduate, hora2021googling, brandt2009two}, indicating the multifaceted approach programmers take towards online learning.

The study conducted by \citet{escobar2019survey}, closely related to our research, surveyed 205 computer science students and developers to understand their learning preferences. Participants were asked to indicate their preferred sources for learning programming topics, including video tutorials, written tutorials, and documentation. The findings showed a preference for combining visual and auditory resources for complex tasks, while straightforward tasks favored written tutorials and documentation.

%Reflect on the previous stuudy 
Our study differs from previous research by focusing on exploring the preferences of developers in utilizing ChatGPT as a learning resource, an area that has not been extensively studied before. Our investigation aims to provide insights into programmer education, complementing existing research on platforms like StackOverflow and YouTube.

\section{Methodology}\label{sec:methodology}
% This section outlines the methodology we used to investigate developers' preferences regarding learning resources in the ChatGPT era. This includes a detailed description of the design of the survey we conducted, details about the survey distribution, and our research questions.

\subsection{Survey Design}
\label{sec:design}
We designed a survey to gather data on developers' preferences for learning resources, with a particular focus on the role that AI chatbots like ChatGPT play in this landscape. The survey began with an informed consent section that included a description of the objectives and scope of the survey, the estimated completion time (\eg 15 minutes), a statement assuring participants of the anonymity of their answers, and provided the contact information of the research team and the Human Research Internal Review Board (IRB) that reviewed and approved the study. The survey was then structured into four main sections, as follows: 1) demographics, 2) preferred learning resources for acquiring software skills and learning new concepts and technologies, 3) preferred learning resources when seeking solutions to programming issues, and 4) the role that AI chatbots play in today's programming learning resources arena. The survey is available in our replication package \cite{replicationPackage}. %, including the perceived advantages, disadvantages, and common uses of AI chatbots.

The first section of our survey gathers basic demographic information from participants, including age, gender, language proficiency, years of software development experience, highest educational degree, current occupation, and country of residence. %This data is critical for analyzing responses across various demographic groups.

In the second section of our survey participants are asked to rank different resources (video tutorials, written tutorials, API documentation, AI chatbots, Q\&A forums, search engines, and others) in order of preference for learning new software development skills or exploring new technologies. This section also inquires about the frequency of AI chatbots' usage as a learning tool.

In the third section, participants rank their go-to resources when looking for solutions to programming issues. The survey includes a variety of resources, similar to the previous section, with the addition of peer discussion platforms (\eg Gitter and Slack). Additionally, the frequency with which participants are using AI chatbots to troubleshoot programming issues is assessed.

The fourth and final section of the survey focuses on the participants' views on the role of AI chatbots in learning programming skills and new technologies, or fixing code issues, as compared to other resources, specifically video tutorials, written tutorials, and Q\&A sites like StackOverflow. First, participants are asked if they believe AI chatbots will replace or complement each of the other types of resources for learning programming-related topics and the reasons behind their beliefs. Then, the survey asks participants about the advantages and disadvantages of AI chatbots they perceive when learning programming topics and finding solutions to programming issues as compared to other programming resources (e.g., video tutorials, written tutorials, and StackOverflow). Finally, the survey concludes with questions about the types of programming-related queries participants typically direct to chatbots, and their assessment of the accuracy and reliability of the answers provided by AI chatbots.

The survey combines multiple choice questions (in sections 1-4 of the survey), ranking questions (in sections 2 and 3), and open-ended questions (in section 4 of the survey) to collect both quantitative and qualitative data. Designed to be completed in approximately 15 minutes, it aims to be thorough, yet not overly burdensome for participants.

% We employed an online survey method, targeting software developers, students, and professionals in software engineering from a variety of demographic backgrounds, experience levels, and geographical areas to achieve a broad participant base. The survey was disseminated across multiple platforms, including social media, developer forums, and Computer Science student mailing lists in various countries to ensure a representative sample.

\subsection{Survey Distribution}
\label{sec:distribution}
This research explores how developers' preferences for educational resources and programming issue solutions have evolved in the ChatGPT era. We distributed the survey using convenience sampling, targeting software developers, computer science students, and faculty. The survey was disseminated across multiple platforms, including social media (\eg X and Facebook), developer groups (\eg on Facebook, Slack, and Discord, with moderator permission), and Computer Science student mailing lists to which the authors had access (\eg Florida State University and University of Jeddah). Since all participants had some programming experience, we refer to our participants as "developers", though not all of them may be currently hired in software development positions. 

We received 341 responses to our survey. However, we excluded 73 of these because they were incomplete responses (\eg participants who did not finish the survey), resulting in a final set of 268 responses used for analysis. 

% We received 341 responses to our survey. However, we had to exclude 73 due to incomplete data, resulting in a final set of 268 responses used for analysis. 

\subsection{Research Questions}
\label{sec:rq}
This research aims to explore and analyze the evolving preferences of developers for learning resources in the context of the ChatGPT era. The questions are designed to understand how AI-driven tools like ChatGPT have influenced the landscape of software development learning and issue solving. The study is structured around the following research questions:

\begin{itemize}
    \item \textbf{RQ$_1$: What resources do developers prefer when learning new software development skills or exploring new technologies?} To answer this research question, we analyze the responses collected from the survey participants in section two of our survey, which asked them to rank various types of resources (e.g., AI chatbots, written tutorials, Q\&A sites, etc.) in order of preference when acquiring new software development skills (RQ$_{1.1}$) and exploring new technologies (RQ$_{1.2}$). The analysis includes calculating the average ranking of the different resources (the metric is described in Section \ref{sec:avg_rank}) and the distribution of respondents' preferences among them.
    \item \textbf{RQ$_2$: What resources do developers consult when faced with programming issues?} To answer this research question, we analyze the responses collected from the survey participants in section three of the survey, which asked them to rank various types of resources in order of preference for programming issue resolution. The analysis includes calculating the average ranking of different resources and the distribution of respondents' preferences among them.
    \item \textbf{RQ$_3$: What is the role of AI chatbots in learning programming topics and finding solutions to programming issues compared to other types of resources?} To answer this research question, we perform quantitative and qualitative analysis of responses to questions in section four of our survey to identify developers' perceptions of AI chatbots as complements or replacements for traditional learning resources such as video tutorials (RQ$_{3.1}$), written tutorials (RQ$_{3.2}$), and Q\&A forums (RQ$_{3.3}$). In addition, we manually analyze through open coding (more details in section \ref{sec:open_coding}) the open-ended answers regarding the perceived advantages and disadvantages of using AI chatbots for learning programming and issue resolution compared to more traditional resources (RQ$_{3.4}$). Lastly, we analyze participants' perceived reliability and accuracy of AI chatbots in providing answers and solutions to programming-related topics and issues (RQ$_{3.5}$).
    \item \textbf{RQ$_4$: How often do developers utilize AI chatbots in their programming learning and issue solving activities?} To answer this research question, we analyze the responses collected in section two of the survey regarding the frequency of use of AI chatbots for learning programming topics and new technologies and for resolving programming issues.
    \item \textbf{RQ$_5$: What are the common programming-related queries or tasks for which survey participants utilize AI chatbots?} To answer this research question, we perform open coding on the open-ended responses to the question in section four of the survey that asks participants about the most common programming-related queries or tasks for which they utilize AI chatbots. 
\end{itemize}

These research questions aim to provide a thorough understanding of the current trends and preferences in learning resources among developers, focusing on the role of AI chatbots in the developers' learning resources.

\subsection{Average Rankings}
\label{sec:avg_rank}
We utilized average rankings to assess developers' preferences for various learning resources in sections two and three of our survey. Participants ranked these resources, with lower rank numbers indicating a higher preference for a resource. To calculate the average ranking across all participants for each resource, we employed the following formula:

\[
\text{Average Rank} = \frac{\sum_{i=1}^{n} \text{Rank}_i}{n}
\]

where $\sum_{i=1}^{n} \text{Rank}_i$ represents the sum of all ranks given to a specific resource by the respondents, and $n$ is the total number of respondents who provided a rank for that resource. This calculation results in an average score that positions each resource relative to the others based on the overall preferences of the participants. A lower average rank indicates a higher preference for that resource.

\subsection{Open Coding}
\label{sec:open_coding}
We employed an open coding process to analyze the open-ended responses from our survey. Open coding is a qualitative data analysis technique that involves breaking down the text into meaningful units, assigning codes to these units, and then categorizing and organizing the coded units into themes or categories \cite{miles1994qualitative,glaser1968discovery,sahar2021issue,diaz2023applying}. We focused on the 1,520 open-ended responses received in six questions, aimed to explore participants' views on AI chatbots like ChatGPT as alternatives or supplements to other software development and maintenance learning resources (i.e., video tutorials, written tutorials, and Q\& sites such as StackOverflow), the perceived advantages and disadvantages of AI Chatbots for programming, and the types of programming-related queries addressed to AI chatbots.

Two annotators conducted the coding independently, reviewing each response for relevant information, identifying meaningful units within the text, and assigning appropriate codes to these units. Given the nature of open-ended questions, multiple codes could be assigned to a single response, allowing for the capture of various perspectives and themes. Following the initial coding phase, annotators convened with a mediator to consolidate similar labels and resolve any discrepancies in their interpretations of the answers. 

Throughout this process, 192 of the 1,520 responses were excluded due to their lack of meaningful content or lack of clarity. After mediation, we calculated Krippendorff's Alpha to assess the reliability of the coding, achieving a score of 0.87 \cite{krippendorff2016reliability,krippendorff2018content}. This indicates a high level of inter-annotator agreement, underscoring the coding process's reliability and the qualitative data's integrity \cite{marzi2024k}.

Analyzing these responses was very labor-intensive, taking a total of 219 work hours to complete. This thorough analysis has yielded valuable insights into developers' nuanced perspectives on the role of AI chatbots in learning programming and issue solving.

\section{Results and Discussion}\label{sec:results}
% \usepackage{multirow}
% \usepackage{array}
% \usepackage{makecell} % For \Cline
% \renewcommand\theadfont{\bfseries} % Optional: To make table head font bold

% Define a new column type "I" for thick vertical lines
% \newcolumntype{I}{!{\vrule width 1.5pt}}

\begin{table*}[]
\centering
\caption{An overview of participants' preferences for various programming learning resources, segmented into three key areas: LS (Learning new Software development skills or concepts), LT (Learning new Technology in software development), and IR (Issue Resolution in programming). It outlines both the top-5 distribution of preferences (\%) and the average ranking for each resource type.}
\label{tab:my-table}
\begin{tabular}{|cIc|c|cIc|c|cIc|c|cIc|c|cIc|c|cIc|c|c|}
\hline
\multirow{2}{*}{Resource/Rank} & \multicolumn{3}{cI}{Rank 1 (\%)} & \multicolumn{3}{cI}{Rank 2 (\%)} & \multicolumn{3}{cI}{Rank 3 (\%)} & \multicolumn{3}{cI}{Rank 4 (\%)} & \multicolumn{3}{cI}{Rank 5 (\%)} & \multicolumn{3}{c|}{Average Rank} \\ \cline{2-19} 
 & LS & LT & IR & LS & LT & IR & LS & LT & IR & LS & LT & IR & LS & LT & IR & LS & LT & IR \\ \Xhline{1.0pt} % Bold bottom border under the header
Video Tutorials & \textbf{53} & \textbf{58} & 13 & 19 & 16 & 17 & 12 & 10 & 16 & 8 & 7 & \textbf{29} & 6 & 5 & 15 & \textbf{2.01} & \textbf{1.96} & 3.49 \\ \hline
Written Tutorials & 10 & 8 & 4 & 20 & \textbf{31} & 9 & 17 & 18 & 18 & \textbf{21} & 20 & 23 & 18 & 13 & \textbf{28} & 3.60 & 3.34 & 4.14 \\ \hline
API Documentation & 6 & 7 & - & 10 & 13 & - & 11 & 20 & - & 15 & 18 & - & \textbf{25} & 16 & - & 4.47 & 4.03 & - \\ \hline
AI Chatbots & 13 & 13 & 30 & \textbf{24} & 18 & 19 & 20 & \textbf{23} & 18 & 16 & 17 & 13 & 12 & 16 & 9 & 3.38 & 3.51 & 2.88 \\ \hline
Q\&A Forums & 4 & 3 & 15 & 6 & 7 & \textbf{31} & 20 & 14 & \textbf{21} & 20 & \textbf{22} & 13 & 23 & \textbf{29} & 14 & 4.34 & 4.47 & 3.01 \\ \hline
Search Engines & 12 & 11 & \textbf{36} & 21 & 16 & 18 & \textbf{21} & 15 & 15 & 20 & 17 & 9 & 13 & 19 & 7 & 3.41 & 3.86 & \textbf{2.79}\\ \hline
Peer Discussion & - & - & 2 & - & - & 6 & - & - & 13 & - & - & 13 & - & - & 25 & - & - & 4.80 \\ \hline
Other & 1 & 0 & 0 & 0 & 0 & 0 & 0 & 1 & 0 & 1 & 0 & 0 & 3 & 2 & 1 & 6.79 & 6.83 & 6.90 \\ \hline
\end{tabular}
\end{table*}

\subsection{Participant Demographics}
In this section, we describe the demographic characteristics of the participants in our survey after excluding incomplete responses, resulting in a final sample size of 268 participants.

% In this section, we describe the demographic characteristics of the participants in our survey after excluding incomplete responses, resulting in a final sample size of 268 participants. The demographic information we collected includes age, gender, native language, educational background, occupation, programming experience, and countries of residence.

% Summary
In our study, 84\% of the participants are young adults (18-24 years), followed by adults aged 25-34 at 11\%. Notably, males represent 75\% of the survey's respondents, a ratio that is representative of the gender disparity commonly found in the software developer population. Our participant pool largely comprises native English and Arabic speakers, at 63\% and 26\% respectively, with a majority residing in the United States (70\%) and Saudi Arabia (26\%). Educationally, the bulk of respondents have a Bachelor's degree (44\%) or a high school diploma (41\%), with a significant portion of the sample being undergraduate and graduate students in Computer Science (57\% and 29\%, respectively). Professional software developers represent another category, with 25 participants or roughly 9\% of our participants. 

Regarding programming experience, 65\% of our respondents are novice programmers with less than 2 years of experience, while those with 2-5 years of experience make up the second largest group at 29\%. The rest of the participants (6\%) have more than 5 years of programming experience.%, as shown in Fig~\ref{fig:experience}. 
This suggests that our participant base has a novice to intermediate level of programming expertise, reflecting a significant number of individuals at the early stages of their programming journey.

\subsection{RQ$_{1.1}$ Preferred Resources for Acquiring New Software Development Skills}
This subsection presents the findings on the preferred resources software developers utilize to acquire new programming skills or concepts. The evaluation is based on both the average ranking assigned to each resource and the distribution of preferences among the survey respondents, as shown in Table~\ref{tab:my-table}.

\subsubsection{\textbf{Average Rankings}}
The analysis commenced with calculating average ranks for each educational resource, determining their positional standings based on the participants' preferences. The resource receiving the lowest average rank score, indicative of higher preference, was video tutorials, with an average rank of 2.01, thus securing the first position. This was closely followed by AI chatbots and search engines, with average ranks of 3.38 and 3.41, respectively, marking them as the second and third preferred resources developers use when trying to learn new programming skills or concepts. Subsequent positions were occupied by written tutorials (average rank of 3.60), Q\&A forums (4.34), and API Documentation (4.47). The 'Other' category of resources, which was offered as a way to refer to any other resources not specifically mentioned in our list, was significantly less preferred, as evidenced by its highest average rank of 6.79, placing it in the final position.

\subsubsection{\textbf{Distribution of Preferences}}

A deeper breakdown of the distribution of rankings across different resources (see Table~\ref{tab:my-table}, columns marked "LS") revealed nuanced insights into the participants' preferences. Video tutorials emerged as the unequivocal favorite among participants, with a majority of respondents (53\%) ranking it as their primary choice. This preference underscores the value placed on visual and demonstrative learning methods. Written tutorials demonstrated a more uniform distribution across the ranks, with the highest concentration (21\%) in the fourth rank, suggesting a moderate preference. API documentation was predominantly ranked towards the lower end of the preference spectrum, with 29\% of respondents placing it in the sixth rank, indicating its use as a more specialized reference tool.

AI chatbots displayed a balanced preference across middle ranks, peaking at 24\% for the second rank. This distribution highlights the growing acceptance of AI chatbots as a supportive educational tool. Q\&A forums and search engines were identified as supplementary resources, with their highest percentages of preference (24\% and 21\%, respectively) falling in the middle to lower ranks. The 'Other' category was overwhelmingly ranked in the last position by 89\% of participants, suggesting a broad consensus on the preference of the specified resources over alternative options.

% A further breakdown into the percentage distribution of rankings across different resources revealed nuanced insights into the participants' preferences:
% \begin{itemize}
%     \item Video tutorials emerged as the unequivocal favorite among participants, with a majority (53\%) ranking it as their primary choice. This preference underscores the value placed on visual and demonstrative learning methods.
%     \item Written tutorials demonstrated a more uniform distribution across the ranks, with the highest concentration (21\%) in the fourth rank, suggesting a moderate preference.
%     \item API documentation was predominantly ranked towards the lower end of the preference spectrum, with 29\% of respondents placing it in the sixth rank, indicating its use as a more specialized reference tool.
%     \item AI chatbots displayed a balanced preference across middle ranks, peaking at 24\% for the second rank. This distribution highlights the growing acceptance of AI chatbots as a supportive educational tool.
%     \item Q/A forums and search engines were identified as supplementary resources, with their highest percentages of preference (24\% and 21\%, respectively) falling in the mid to lower ranks.
%     \item Remarkably, the 'Other' category was overwhelmingly ranked in the last position by 89\% of participants, suggesting a broad consensus on the preferability of the specified resources over alternative options.
% \end{itemize}

The findings highlight a pronounced preference for video tutorials among participants seeking to acquire new programming skills and concepts, likely attributed to the format's ability to offer an engaging, multimedia experience and practically demonstrative content. AI chatbots and search engines were also identified as significant, reflecting a trend towards interactive and easily navigable information sources. Although written tutorials, Q\%A forums and API documentation remain valuable, they have varied rankings. From information we learned while analyzing open-ended answers for other research questions, we hypothesize that the learner's specific requirements, the nature of the information sought, and personal learning styles may influence their utility in specific situations. %The negligible preference for 'Other' resources further validates the effectiveness and popularity of the identified resources within the software development community.

\subsection{RQ$_{1.2}$ Preferred Resources for Exploring New Technologies}
This subsection analyzes the resource preferences of developers when exploring new technologies, as indicated by the average ranking scores and the preference distribution among the surveyed population (Table~\ref{tab:my-table}).

\subsubsection{\textbf{Average Rankings}}
The investigation into the preferred learning resources for exploring new technologies mirrors a similar pattern as the one observed in RQ1.1. Video tutorials again rank as the most favored resource, with an average rank of 1.96, affirming their top position. Written tutorials occupy the second slot with an average rank of 3.34, followed by AI chatbots at 3.51 and search engines at 3.86, placing them in third and fourth positions, respectively. API documentation and Q\&A Forums are further down the list, with average ranks of 4.03 and 4.47, respectively. The 'Other' category remains the least preferred, with an average rank of 6.83.

\subsubsection{\textbf{Distribution of Preferences}}

The percentage distribution across ranks for each resource offers additional insights (see Table~\ref{tab:my-table}, columns marked "LT"). Video tutorials continue to dominate, with 58\% of respondents ranking them as their primary choice, emphasizing the predominant preference for visual/auditory learning methods in the exploration of new technologies. Written tutorials also show a substantial preference, particularly in the second rank, where 31\% of participants placed them, indicating their value in providing in-depth explanations.

API documentation is also recognized as important, with a steady distribution across the middle ranks, peaking at 25\% in the sixth rank. AI chatbots have a strong presence in the middle ranks, with the highest concentration (23\%) in the third rank, which indicates their growing importance as a learning aid.

Q\%A forums and search engines show a more dispersed preference across the ranks, with a notable portion of respondents (29\% and 22\%, respectively) ranking them in the lower positions. The 'Other' category, was overwhelmingly ranked last by 90\% of respondents, indicating a clear preference for the specified resources over less common alternatives when exploring new technologies.

% The percentage distribution across ranks for each resource offers additional insights:
% \begin{itemize}
%     \item Video Tutorials continue to dominate, with a remarkable 58\% of respondents ranking them as their primary choice, emphasizing the prevalent preference for visual learning methods in the exploration of new technologies.
%     \item Written Tutorials show a substantial preference as well, particularly in the second rank, where 31\% of participants placed them, indicating their value in providing in-depth explanations.
%     \item API Documentation is recognized for its importance, with a steady distribution across the middle ranks, peaking at 25\% in the sixth rank. This suggests its role as a detailed reference point in the learning process.
%     \item AI Chatbots exhibit a strong presence in the middle ranks, with the highest concentration (23\%) in the third rank, signifying their growing significance as an interactive learning aid.
%     \item Q/A Forums and Search Engines show a more dispersed preference across the ranks, with a notable portion of respondents (29\% and 22\%, respectively) ranking them in the lower positions, highlighting their use as supplementary resources.
%     \item The 'Other' category, overwhelmingly ranked last by 90\% of respondents, indicates a clear preference for the specified resources over less common alternatives when exploring new technologies.
% \end{itemize}

The preference for video tutorials when exploring new technologies again highlights the critical role of auditory-visual and demonstrative learning. Written tutorials and AI chatbots also play a significant role, suggesting that developers may value a blend of in-depth materials and interactive engagement. Search engines and Q\&A forums, despite being ranked lower, are indispensable for quick searches and community-driven insights. API documentation's position reflects its necessity for detailed technical understanding. The marginal preference for 'Other' resources highlights the effectiveness of these conventional resources in catering to developers' needs in exploring new technologies, reinforcing the observed trends in acquiring new software development skills.

\subsection{RQ$_2$ Resources Consulted for Programming Issues and Errors}
This subsection investigates the resources to which developers turn when faced with programming problems and errors. The analysis includes the average ranking of each resource and their distribution of preferences as indicated by the survey responses (Table~\ref{tab:my-table}).

\subsubsection{\textbf{Average Rankings}}
Participants exhibit a slightly different pattern of resource preference when resolving programming issues and errors compared to acquiring new skills or exploring new technologies. Search engines emerge as the most consulted resource, with an average rank of 2.74, earning them the top position. Following closely are Q\&A forums and AI chatbots, with average ranks of 2.90 and 2.99, respectively, indicating their critical roles in issue resolution. Video tutorials, written tutorials, and peer discussions (i.e., Slack, Discord, Gitter, etc.) are ranked in the subsequent positions with average ranks of 3.60, 4.11, and 4.77, respectively. The 'Other' category of resources remains the least consulted, with an average rank of 6.89.

\subsubsection{\textbf{Distribution of Preferences}}

The distribution of preferences across ranks for each resource (see Table~\ref{tab:my-table}, columns marked "IR") reveals specific tendencies in consulting resources for programming issues. Search engines stand out, with 36\% of respondents ranking them as their primary go-to resource. AI chatbots are also a popular resource for issue resolution, with 30\% of participants ranking them as their top choice, reflecting the emerging reliance on AI for immediate and interactive issue resolution. Q\%A Forums are also highly favored, with 31\% of participants ranking them second, showcasing the value of community-driven problem solving advice and shared experiences.

Video tutorials and written tutorials, while not in the top, still demonstrate their continued usefulness, with participants' preferences for these resources being spread across various ranks, peaking at 29\% and 28\% in the fourth and fifth ranks, respectively. Peer discussion is dominant in the later ranks, with 40\% of participants placing it sixth, indicating its role as a supplementary resource rather than a primary source of solutions.

The 'Other' category, overwhelmingly ranked last by 93\% of respondents, underscores a strong consensus on the utility of the mentioned resources over less conventional alternatives for troubleshooting programming issues.

% The distribution of preferences across ranks for each resource reveals specific tendencies in consulting resources for programming issues:

% \begin{itemize}
%     \item Search Engines stand out, with 36\% of respondents ranking them as their primary go-to resource, emphasizing the importance of quick, accessible solutions.
%     \item Q/A Forums are also highly favored, with 31\% of developers ranking them first, showcasing the value of community-driven advice and shared experiences.
%     \item AI Chatbots have seen a notable increase in use, with 30\% ranking them as their top choice, reflecting the emerging reliance on AI for immediate, interactive issue resolution.
%     \item Video Tutorials and Written Tutorials demonstrate their continued importance, with a preference spread across the ranks, peaking at 29\% and 28\% in the fourth and fifth rank, respectively, suggesting their use for more in-depth understanding of solutions.
%     \item Peer Discussion is highlighted in the later ranks, with 40\% placing it sixth, indicating its role as a supplementary resource rather than a primary source of solutions.
%     \item The 'Other' category, overwhelmingly ranked last by 93\% of respondents, underscores a strong consensus on the utility of the mentioned resources over less conventional alternatives for troubleshooting programming issues.
% \end{itemize}

The findings illustrate a clear preference for search engines and Q\&A forums as primary resources for addressing programming issues and errors, underscoring the usefulness of searchable solutions and the collective knowledge of the programming community. AI chatbots were also a popular resource, reflecting the evolving landscape of developer tools and an interest in the immediate and personalized assistance these tools offer. The results also indicate that video and written tutorials still serve an essential role for comprehensive learning and understanding, while peer discussion can also be useful, likely due to the community support they offer.

% \begin{figure}[]
% \centering
% \includegraphics[width=1\linewidth]{Figures/frequency_of_use.pdf}
% \caption{Comparative frequencies of using AI chatbots for programming learning and issue resolution among survey participants}
% \label{fig:frequency_of_use}
% \end{figure}

\subsection{RQ$_{3.1}$  AI Chatbots - Replacing or Complementing Video Tutorials for Learning Programming-Related Topics?}
% In exploring the potential impact of AI chatbots on the utilization of video tutorials for learning programming-related topics, our survey responses reveal various insights. 
A vast majority of participants, 212 out of 268 or approximately 79\%, view AI chatbots as a complement to video tutorials rather than a replacement, as shown in Fig~\ref{fig:role_of_ai}.  Meanwhile, a smaller segment of the respondents, 45 individuals or about 17\%, believe that AI chatbots could replace video tutorials as a learning resource for programming concepts. Only a very small group of 11 participants or roughly 4\%, chose a third option available, "Other", which allowed them to then fill in their own answer to describe the relationship between AI chatbots and video tutorials, in case their opinion did not fit neatly into neither the "Replace" nor the "Complement" options available.  These findings suggest confidence in the growing capabilities of AI chatbots to deliver educational content effectively, although they also indicate that these tools are not ready to fully replace programming video tutorials as a learning resource. These findings underscore the perceived value of integrating AI chatbots with traditional video tutorials, suggesting that a hybrid approach may offer the most benefits for learners in programming-related fields. 

In our qualitative analysis of the reasoning provided by participants regarding their opinion about AI chatbots' role with respect to video tutorials, we identified several main themes reflecting participants' nuanced perspectives. For participants whose responses indicated that AI chatbots complement video tutorials, the reasons provided included advantages of video tutorials over AI, such as detailed explanations provided by humans, ease of following the presented material, the ability to visualize concepts effectively, and the ability to accommodate different learning styles. 

For instance, one participant stated, \textit{"Video explanations are usually more thorough, and using the AI bot to complement it, we might be able to understand the subject explained in the video better. But I suppose AI bots are unable to generate a completely new instruction that nobody has mentioned before so replacing it wouldn't be very possible."} Another participant commented, \textit{"I think that video tutorials are helpful for step by step visual instruction, and visual instruction isn't something that AI chatbots can replace, but they can be helpful to clarify steps from the video."}

In contrast, our examination of responses indicating AI replacing video tutorials identified advantages and capabilities of AI chatbots, highlighting the personalized learning experiences they offer, and their ability to provide faster and more convenient assistance. For example, one participant said, \textit{"I feel that AI chatbots will replace video tutorials solely due to their interactive nature. If there is a concept you don't understand from what the AI chatbot initially told you, for example, then you can ask for elaboration, clarification, or ask it to completely start over and deliver the information in a different manner. Video tutorials don't allow for such dynamic interaction."}

An analysis of the reasonings provided for the 'Other' category revealed some new topics such as concern for low-quality video tutorials being created using AI, but also identified some themes brought up in the "Replace" and "Complement" answers, such as identifying advantages of both types of resources. For example, one participant commented \textit{"Most videos available today are driven by monetary objectives, including tutorial videos. I feel it will be very hard to distinguish quality video tutorials, because AI-generated content is designed to appeal to human liking. Also, AI will allow inexperienced individuals to make and publish video tutorials for the sake of 'making content,' which can attract and fool beginners."} We include these results, as well as more detailed information on the analysis in our replication package \cite{replicationPackage}.

\subsection{RQ$_{3.2}$ AI Chatbots -  Replacing or Complementing Written Tutorials for Learning Programming-Related Topics?}
% The survey results reveal participants' perspectives on the impact of AI chatbots on traditional written tutorials for learning programming-related topics. 
A significant portion, 154 individuals or approximately 57\% of the respondents, believe that AI chatbots can replace written tutorials, as shown in Fig~\ref{fig:role_of_ai}. This contrasts with the earlier finding where AI chatbots were predominantly seen as supplements to video tutorials, suggesting a nuanced view of the role of AI chatbots depending on the format of the learning resource. On the other hand, 110 participants, making up about 41\% of the sample, view AI chatbots as complementary to written tutorials, indicating that both resources can coexist beneficially to enhance the learning experience. A small group of 4 individuals, or roughly 1.5\%, provided other insights. These observations highlight a nuanced perspective among learners, indicating a preference for AI chatbots as complement to video resources while presenting a stronger disposition toward replacing written tutorials. This juxtaposition suggests that the format of the learning resource (video vs. written) may influence how AI chatbots are perceived and utilized in educational contexts for programming-related topics.

In our qualitative analysis of responses concerning AI chatbots' impact on written tutorials as programming-related learning resources, we identified diverse themes. For those viewing AI chatbots as replacements for written tutorials, their reasoning mentioned advantages and capabilities of AI chatbots, highlighting their ability to provide personalized, interactive, and fast responses, including step-by-step, detailed explanations and answers tailored to specific questions or issues. One participant stated, \textit{"It could replace written tutorials as AI chatbots and written tutorials both provide information via a written layout. AI chatbots hold an advantage in terms of providing more information tailored to the user's questions that not all written tutorials may answer."}

% \textit{"I think AI could mostly replace written tutorials since the user could prompt the chatbots to provide the user with written tutorials. So, seeing as AI chatbots can provide the user with written tutorials and can be manipulated to provide further information on certain topics, I don't see a need for written tutorials outside of AI chatbots (other than written tutorials specifically FOR the chatbots)."}

In contrast, our examination of responses favoring the complementary role of AI chatbots revealed appreciation for both AI chatbots' speed and personalization and written tutorials' clarity and depth. For example, one participant commented, \textit{"While AI chatbots will provide a solution to a problem quicker, written documentation offers a better understanding of the problem to the reader."}

% \TODO{the quotes you provide should exemplify what you just said above as a general note. This example does not exemplify the "clarity and depth" or written tutorials, nor the "speed and personalization" of AI chatbots. Please check that all examples exemplify your main message, otherwise they are confusing the reader. } 
% For example, one participant commented \textit{"I think being written by a human provides a huge advantage. They've been in our shoes, and I don't think machine learning is at the point yet where it can correctly help with all programming-related topics. There are always experts that exist in each topic. Chatbots can use to supplement the written material." }

An additional analysis of the 'Other' category revealed an emphasis on AI chatbots' efficiency and the continuous relevance of traditional educational approaches. For instance, one participant mentioned, \textit{"The AI chatbot or copilot saves the developer time for getting the information, but for beginners, they need full course sessions to understand the basics."} Due to space constraints, we provide a more detailed analysis in the replication package \cite{replicationPackage}.

\subsection{RQ$_{3.3}$ AI Chatbots - Replacing or Complementing Q\&A Forums for Learning Programming-Related Topics?}
\begin{figure}[]
\centering
\includegraphics[width=1\linewidth]{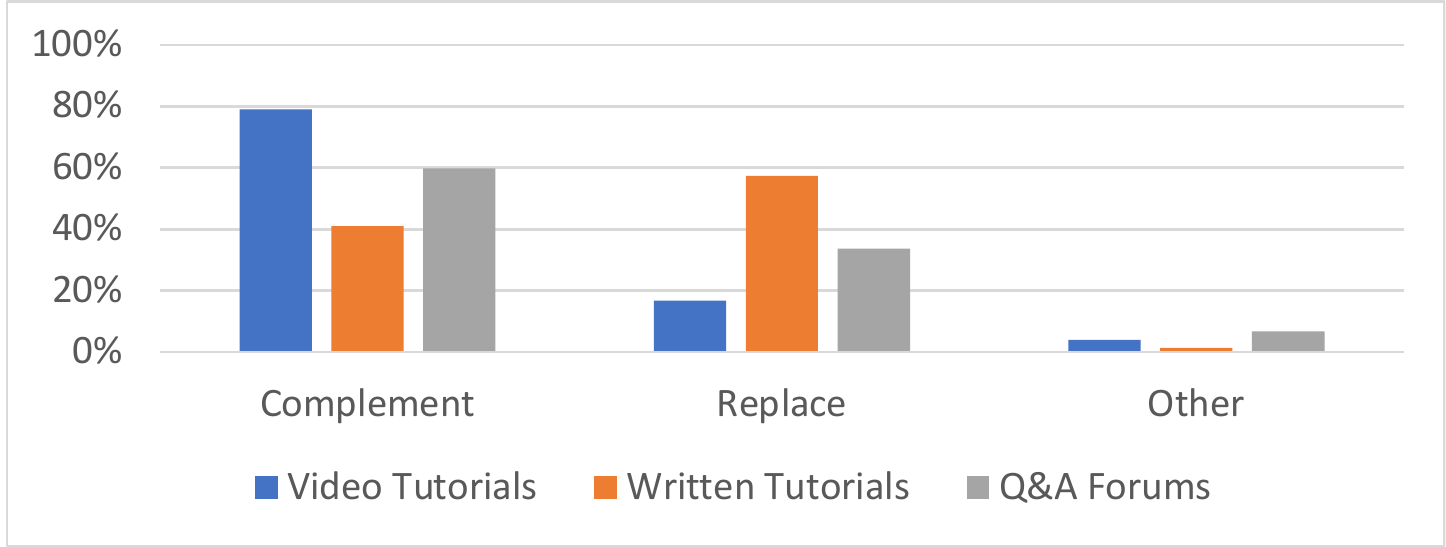}
\caption{Perceptions of AI Chatbots' Role with Respect to Other Programming Learning Resources}
% \caption{Perceptions of AI Chatbots' Role in Programming Learning Resources among survey participants}
\label{fig:role_of_ai}
\end{figure}

% The survey data offers insight into how participants perceive the role of AI chatbots in relation to Q\&A forums for learning programming-related topics.
Most respondents, 160 or approximately 60\%, view AI chatbots as a complement to Q\&A forums, as shown in Fig~\ref{fig:role_of_ai}. This suggests recognizing the unique benefits that both AI chatbots and Q\&A forums can provide. Meanwhile, 90 participants, about 34\% of the sample, believe that AI chatbots could replace Q\&A forums. This significant portion indicates confidence in AI chatbots' ability to offer comprehensive and accurate answers, potentially streamlining the search for solutions. A smaller group, 18 individuals or roughly 7\%, shared other views, reflecting a range of opinions on the integration and utility of AI chatbots alongside traditional Q\&A forums. These findings suggest a balanced perspective among learners, with a lean towards enhancing existing Q\&A forums with AI chatbots rather than viewing them as outright replacements, especially in contexts where human interaction and community feedback are valued.

In the qualitative analysis of the responses to the open questions asking participants to provide a motivation for their response, we identified multiple themes for those viewing AI chatbots as complements to Q\&A forums. These themes emphasized some advantages of Q\&A forums, such as the variety of experiences and solutions shared in them, the benefits of the presence of human interaction, having a sense of community, and a better understanding and relatability of answers provided. They also emphasized some benefits of AI chatbots, like immediate, interactive, and personalized responses. 

For example, one participant stated, \textit{"AI chatbots are helpful for giving interactive help and solutions, but Q\&A forums are still helpful for the variety of experiences and solutions that other people come up with, as well as online communities being helpful."} Another participant pointed out, \textit{"It will reduce the amount of questions on the Q\&A sites. But sometimes people prefer to ask others who fall in the same problem before so that the answer would be trusted."}

Among participants considering AI chatbots as replacements for Q\&A forums, the themes focused on the benefits and advantages of AI chatbots, such as speed, personalized assistance, and the ability to enhance issue resolution through immediate feedback. One participant commented, \textit{"I find that AI chatbots are more efficient at providing help than Q\&A Forums. With Q\&A forums, you have to wait for someone to respond to your question if it has not already been answered and there is no guarantee that the response will be helpful."}

Lastly, when participants chose the 'Other' category, analyzing the responses actually revealed themes related to the value of Q\&A forums, highlighting their irreplaceable role in offering depth and diverse perspectives. For instance, a participant stated, \textit{"I don't think AI chatbots will ever replace Q\&A forums, because on Q\&A forums real experiences of developers are shared; basically real communication happens with experienced developers, and that's why I think this way."} Due to page limit constraints, this summary briefly outlines the thematic findings; detailed insights are available in the replication package \cite{replicationPackage}.

\subsection{RQ$_{3.4}$: Advantages of Using AI Chatbots for Learning Programming Topics and Finding Solutions Compared to Other Resources}
The analysis of participants' views on the advantages of using AI chatbots over traditional programming resources reveals several key themes emphasizing the unique benefits that AI chatbots offer in learning programming and solving programming-related issues. We excluded 11 responses from our analysis because they were empty or unintelligible, therefore focusing our analysis on the remaining 257 valid responses. The themes identified span across efficiency, accessibility, personalization, educational support, issue resolution capabilities, information management, and interactive feedback.

% \begin{figure}[]
% \centering
% \includegraphics[width=1\linewidth]{Figures/wordcloud_advantages.png}
% \caption{Word Cloud of Participants' Perceptions of Advantages of AI Chatbots Compared to Other Programming Learning Resources}
% \label{fig:wordcloud_advantages}
% \end{figure}

\textbf{Efficiency, Accessibility, and Convenience (221 mentions):} This theme emerged as the most pronounced, with participants highlighting the speed and ease of access to information provided by AI chatbots. The ability to receive direct, accurate, and concise answers anytime and anywhere highlights the high value placed on AI chatbots' efficiency and user-friendly nature. Their flexibility, availability, and convenience make them an appealing alternative to traditional, often more time-consuming, resources.

\textbf{Personalization and Adaptability (84 mentions):} Participants appreciated the ability of AI chatbots to deliver personalized and specific answers to queries. This adaptability, including tailored assistance and the capacity to understand and adjust to user feedback, suggests a significant advantage in using AI chatbots for learning and issue resolution.

\textbf{Learning and Educational Support (75 mentions):} The role of AI chatbots in offering clear explanations and more examples, interactive assistance, and acting as on-demand tutors highlights their contribution to a supportive and engaging learning environment. Participants valued the non-judgmental interaction and the ability of chatbots to facilitate understanding, suggesting that AI chatbots can effectively supplement traditional educational resources.

\textbf{Issue Resolution and Technical Assistance (50 mentions):} AI chatbots' practical assistance in coding and debugging, providing specific solutions, and enhancing the quality of work illustrates their technical value. Their ability to generate boilerplate code, write test cases, and offer dynamic troubleshooting emphasizes the practical, hands-on benefits of using AI chatbots for programming tasks.

\textbf{Information Management and Summarization (26 mentions)}: AI chatbots' ability to provide summaries and gather information from a wide range of sources is seen as a crucial advantage. This capability, along with providing all-in-one solutions and access to vast data sets, underscores the role of AI chatbots in managing and synthesizing information efficiently.

\textbf{Interaction and Feedback (10 mentions):} The interactive nature of AI chatbots, allowing for follow-up questions and clarifications, was also highlighted. This feedback mechanism and the ability to refine user queries through interaction add a layer of engagement and learning opportunities not always present in other resources.

\textbf{Additional Specific Capabilities (4 mentions): }Unique benefits such as identifying key concepts, suggesting areas for knowledge improvement, and suitability for straightforward tasks reveal the diverse functionalities of AI chatbots that participants find valuable.

In summary, participants see AI chatbots as highly beneficial for learning programming due to their efficiency, accessibility, personalized support, and interactive feedback. These chatbots are valued for their ability to provide quick, tailored assistance, facilitate deeper learning, and offer practical issue resolution support, distinguishing them from traditional programming resources.

\subsection{RQ$_{3.5}$: Disadvantages of Using AI Chatbots for Learning Programming Topics and Finding Solutions Compared to Other Resources}
The analysis of participants' views on the disadvantages of using AI chatbots to learn programming topics and find solutions compared to traditional resources reveals concerns across multiple themes. We excluded 38 responses from our analysis because they were empty, did not provide answers to the questions asked, or were unintelligible. Our analysis then focused on the remaining 230 valid responses. These themes reflect concerns about reliability, accuracy, dependency, depth of the learning experience, technical limitations, ethical considerations, interaction quality, and the dynamics of human learning.

% \begin{figure}[]
% \centering
% \includegraphics[width=1\linewidth]{Figures/wordcloud_disadvantages.png}
% \caption{Word Cloud of Participants' Perceptions of Disadvantages of AI Chatbots Compared to Other Programming Learning Resources}
% \label{fig:wordcloud_disadvantages}
% \end{figure}

\textbf{Concerns About Reliability and Accuracy (180 mentions):} The most significant concern among participants revolves around the reliability, accuracy, and consistency of AI chatbots. Common issues include misunderstanding or misinterpreting questions, fabricating plausible yet incorrect information, and providing solutions that may not be optimal or efficient. Concerns also extend to AI chatbots generating overly complex code, offering incomplete answers, and sometimes failing to explain the issue resolution process adequately. The need to verify AI-provided solutions and the possibility of outdated solutions amplify worries about the dependability of AI chatbots as learning and issue resolution tools.
    
\textbf{Dependence and Overreliance Issues (62 mentions):} Participants expressed worries about the potential for AI chatbots to foster dependence or overreliance, potentially hindering the development of independent research, learning, and problem-solving skills. Concerns were also raised about diminishing human interactions, community engagement, creative thinking, and critical thinking skills due to reliance on AI for quick solutions.

\textbf{Technical and Specific Limitations (34 mentions):} Technical limitations, such as the lack of visual aids or demonstrations and the challenge of conveying complex visual concepts, were highlighted. Other concerns include verbose responses, limited assistance for advanced topics, and challenges in addressing project-specific constraints or providing contextually rich answers. The effectiveness of AI chatbots also seems contingent on the user's ability to ask the right questions, potentially limiting their utility for those less familiar with the subject matter.

\textbf{Usage and Interaction Challenges (21 mentions):} The effectiveness of AI chatbot responses was seen to heavily depend on users' ability to craft precise prompts, with prior knowledge of the topic often necessary for understanding explanations. Generalized responses, technical language, and challenges in guiding beginners effectively were also noted as barriers to effective use.

\textbf{Learning and Understanding Challenges (18 mentions):} The potential for AI chatbots to encourage superficial information gathering over deep understanding was noted, with concerns about deterring in-depth comprehension of concepts and diminishing incentives for deep thinking and comprehensive understanding.

\textbf{Ethical and Resource Concerns (9 mentions):} Ethical considerations, resource intensity of AI operations, and potential impacts on human resources were among the concerns raised, pointing to broader implications of AI chatbot integration into learning and issue resolution contexts.
    
\textbf{Challenges in Adaptation and Feedback (7 mentions):} The limited ability of AI chatbots to adjust solutions based on feedback and their lack of critical thinking abilities were highlighted as drawbacks, alongside concerns about the credibility and real-life applicability of AI-provided solutions.
    
\textbf{Human Interaction and Learning Dynamics (3 mentions):} A preference for human explanations and interactions over AI-provided information was expressed, along with concerns about chatbots providing excessive or unnecessary information that could lead to confusion.

In summary, while AI chatbots offer numerous advantages in terms of accessibility, convenience, and personalized learning, these insights underscore significant reservations regarding their reliability, depth of learning, technical limitations, and the potential for fostering an overreliance that could inhibit skill development. These findings suggest a balanced approach, integrating AI chatbots as supplementary tools alongside traditional resources, may mitigate these disadvantages, leveraging the strengths of AI while addressing its limitations.

\begin{figure}[]
\centering
\includegraphics[width=1\linewidth]{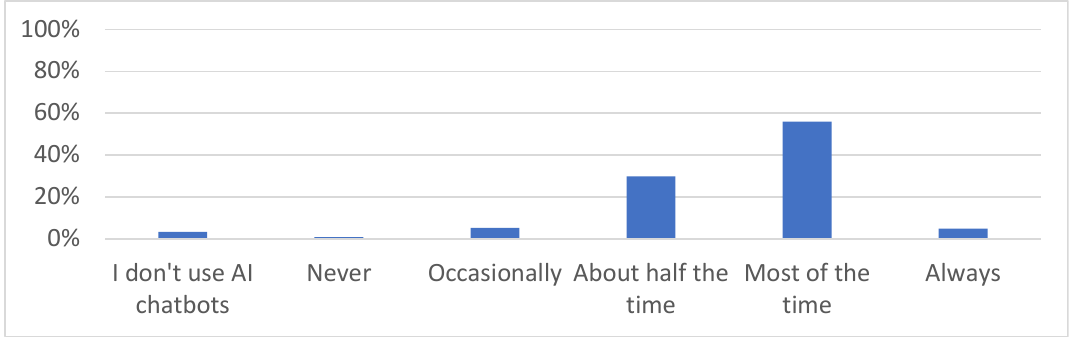}
\caption{Survey responses on the reliability and accuracy of AI chatbots}
\label{fig:reliability_of_ai}
\end{figure}

\subsection{RQ$_{3.6}$: Accuracy and Reliability of Answers or Information Provided by AI Chatbots}
Survey data on the perceived reliability and accuracy of AI chatbots (Fig~\ref{fig:reliability_of_ai}) reveals that a majority, 150 respondents or 56\%, find AI chatbots reliable and accurate 'Most of the time', indicating a high level of trust in these tools for programming assistance. About 80 participants, or 30\%, perceive chatbots as accurate 'About half the time', suggesting mixed experiences with their consistency and accuracy. A smaller group, 14 or 5\%, only occasionally finds AI chatbots reliable, while an optimistic 13 users, or 5\%, report 'Always' receiving accurate information from them. On the other end, a minimal number, 2 or less than 1\%, believes chatbots 'Never' provide reliable answers, and 9 participants, or 3\%, have not used AI chatbots at all.

This distribution reflects general confidence in AI chatbots as useful resources for programming, with a significant majority reporting positive experiences. However, the variation in responses also highlights the presence of skepticism and varied experiences among users and points to areas for improvement in enhancing the reliability and accuracy of AI chatbots to more effectively meet diverse user needs.

\subsection{RQ$_{4}$: Utilization of AI Chatbots in Programming Learning and Problem-Solving Activities}
% \subsection{RQ How often do survey participants utilize AI chatbots in their learning processes and problem-solving activities within software development and programming?}

% Our survey investigated the frequency with which participants use AI chatbots for both learning and problem-solving within the context of software development and programming. The results highlight a significant engagement with AI chatbots among our participants, showcasing diverse usage patterns for both educational purposes and issue resolution.

% , as shown in Fig~\ref{fig:frequency_of_use}
For programming learning, a majority of participants report using AI chatbots on a weekly basis (47\%), followed by daily usage (31\%). This indicates a strong reliance on AI chatbots as a regular resource for acquiring programming knowledge and skills. Additionally, 10\% of participants use chatbots monthly, and another 10\% rarely utilize them for learning. Only a small fraction (3\%) of the sample has never used AI chatbots for this purpose.

In the context of problem-solving or issue resolution within software development, weekly usage of AI chatbots also dominates, reported by 41\% of participants. This is closely followed by daily usage, with 33\% of participants relying on chatbots to assist with troubleshooting and solving programming challenges. Monthly usage is slightly higher for issue resolution (14\%) compared to learning, indicating that some participants may specifically seek out AI chatbots for help with specific problems. Similarly, 8\% of participants rarely use AI chatbots for problem-solving, and a minimal 3\% have never used them for this purpose.

These findings reveal a strong inclination among the survey participants to use AI chatbots regularly and consistently, highlighting their importance as a tool for both educational enhancement and practical support in coding activities.

\subsection{RQ$_{5}$: Common Programming-Related Queries or Tasks for Which Survey Participants Utilize AI Chatbots}
% \subsection{RQ What are the most common types of programming-related queries or tasks for which survey participants utilize AI chatbots?}

The analysis of survey responses, excluding 53 due to being empty, unclear, or not relative to the questions asked, out of 268, has identified several distinct themes reflecting the diverse applications of AI chatbots in programming-related tasks. These themes highlight the versatility of AI chatbots in addressing a broad spectrum of programming needs, from troubleshooting and code development to educational support and practical application.

% The analysis, after excluding 53 responses from a total of 268 for being unclear or irrelevant, identified key themes in the utilization of AI chatbots for programming tasks. These themes span from code troubleshooting to educational support:

\textbf{Learning and Understanding Concepts (156 mentions):} AI chatbots are used for guidance, code explanation, and concept understanding, showcasing their educational benefits.

\textbf{Debugging and Error Resolution (129 mentions):} AI chatbots play a crucial role in debugging, error explanation, and syntax resolution, highlighting their importance in enhancing code quality.

\textbf{Code Development and Enhancement (52 mentions):} AI chatbots assist in code generation, improvement, documentation, and testing, emphasizing their significance in the coding workflow.

\textbf{Practical Application and Task Automation (10 mentions):} AI chatbots have applications in writing scripts, information summarization, and idea generation and improvement.
    
\textbf{Academic Support, Ethics, and Visualization (4 mentions):} AI chatbots also find applications in academic assistance, ethical questions, and data visualization.

In summary, our analysis reveals a wide spectrum of programming-related queries and tasks for which AI chatbots are utilized, with debugging and error resolution, code development and enhancement, and learning and understanding concepts being the most prominent. These findings suggest that AI chatbots are valued not only for their technical assistance but also for their educational support. Due to page limit constraints, this summary briefly outlines the thematic findings; detailed insights are available in the replication package \cite{replicationPackage}.

% \section{Discussion}\label{sec:discussion}
% \input{Discussion.tex}

\section{Threats To Validity}\label{sec:threats_to_validity}
In this study, we identify and address several threats to the validity of our findings, categorizing them into constructs and internal and external validity threats.

Threats to construct validity include the self-reported nature of our data, which may be subject to recall bias or social desirability bias, in which participants may provide responses that they believe align with societal expectations rather than their true preferences or experiences. To address this, we ensured anonymity in data collection, hoping to encourage more honest and accurate responses. Another concern is the possibility of not listing all learning resources that participants may have used in our survey, leading to potential underrepresentation or bias in our findings. To mitigate this, we included 'other' as an option for participants to input any additional learning resources not listed in the survey. However, the results show that most participants put these "other" resources as the least preferred resources in all questions, which suggests that our survey captures the commonly used learning resources.

Threats to internal validity include the potential for subjectivity in interpreting responses during the open coding process. To mitigate this threat, two annotators independently reviewed and labeled the responses, and any discrepancies were resolved through discussion and consensus among the annotators with assistance from a mediator. We also excluded ambiguous responses that could not be reliably coded from our analysis. Additionally, we utilized Krippendorff's alpha to measure inter-coder reliability, and the computed value indicated a high level of agreement between annotators.

External validity threats include the limited generalizability of our findings due to our sample's specific demographics and geographical distribution. The concentration of respondents from specific regions or with similar demographic profiles might have influenced the study outcomes, underscoring the need for future research to engage a broader, more varied population to validate and extend our findings.

% \section{Lessons Learned}
\section{Lessons Learned and Actionable Results}

Our study highlights developers' evolving preferences for learning resources in the era of generative AI, focusing on AI chatbots like ChatGPT. This section discusses the key lessons learned from our findings and offers actionable results for educators and the software development community to improve the learning experience.

\subsection{Lessons Learned}
Participants show a strong preference for video tutorials when acquiring new skills and exploring new technologies, emphasizing the importance of visual and demonstrative content. Video tutorials are highly valued for their ability to visually demonstrate concepts and procedures. In addition, the role of AI chatbots varies depending on the type of learning resource. The majority of respondents view AI chatbots as valuable complements to video tutorials and Q\&A forums, providing instant, personalized assistance. Many respondents also believe that AI chatbots can replace written tutorials due to their ability to provide the same information, but more concise, direct answers. However, reliability and accuracy of AI chatbot responses are major concerns. Instances of chatbots providing incorrect or incomplete information highlight the need for improvement in this area.

\subsection{Actionable Results}
Educators and the software development community should aim to achieve a balanced integration of AI chatbots with traditional learning methods. AI chatbots offer immediate personalized support, while traditional resources such as videos and written tutorials provide in-depth, structured content. An integration of the two types of resources would leverage the strengths of both, creating a comprehensive and effective learning environment. Additionally, to mitigate the risk of overreliance on AI chatbots, which a large proportion of participants reported as a negative aspect, it is important to encourage learners to engage deeply with the material in traditional educational resources. This can help develop independent problem-solving skills and critical thinking abilities.

\section{Acknowledgments}
% This work is supported in part by the National Science Foundation grant CCF-1846142. 
This work is supported in part by the National Science Foundation grant CCF-1846142. Ahmad Tayeb is sponsored in part by the Saudi Arabian Cultural Mission (SACM) and King Abdulaziz University (KAU).

\section{Conclusions and Future Work}
\label{sec:Conclusion}
Our study explored how large language models (LLMs) like ChatGPT influence software developer learning resources. Surveying developers and computer science students, we found from 268 analyzed responses a trend towards interactive learning. Video tutorials are favored for skill acquisition, but AI chatbots are becoming popular as supplemental tools for their personalized learning experiences. Many respondents see AI chatbots as potential replacements for written tutorials, offering concise, direct answers and suggesting a role for chatbots depending on the learning format. Traditional search engines and Q\&A forums remain vital for quick information and community knowledge. The study shows AI chatbots as complements or replacements based on the learning context, with written and video tutorials preferred for in-depth learning. AI chatbots excel in providing immediate, customized explanations. Future research could focus on leveraging AI chatbots to enhance traditional resources and improve their accuracy.

\bibliographystyle{plainnat}
\bibliography{base}

\end{document}